# Nanometer thick magneto-optical iron garnet films

Gribova[1,2] N.I., Osmanov[3] S.V., Lyashko[3] S.D., Shilina[1] P.V., Mikhailova[3] T.V., Polulyakh[3] S.N., Berzhansky[3] V.N., Wang[4] Xianjie, Han[5] Xiufeng, Belotelov[1,6] V.I.

[1]Russian Quantum Center, Moscow 121205, Russia
[2]Moscow Institute of Physics and Technology (National Research University), Dolgoprudny 141701, Russia
[3]V.I. Vernadsky Crimean Federal University, Simferopol 295007, Russia
[4]School of Physics, Harbin Institute of Technology, Harbin 150001, China
[5]Beijing National Laboratory for Condensed Matter Physics, Institute of Physics, Chinese Academy of Sciences, Beijing 100190, China
[6]Lomonosov Moscow State University, Moscow 119991, Russia

**Abstract**

Here we demonstrate nanometer thick iron garnet films suitable for the magneto-optical applications. Bismuth-substituted iron garnet films of compositions $Bi_1Y_2Fe_5O_{12}$ and $Bi_1Tm_2Fe_5O_{12}$ deposited on gadolinium gallium garnet substrate are fabricated and characterized. Their thicknesses range from 2 to 10 nm, which corresponds to just a few crystal lattice constants. Faraday rotation of the nanofilms reaches 29.7 deg/µm at 420 nm which is comparable and even a bit better than single crystal micrometer thick films of similar composition. The film surface morphology by atomic force microscopy gives root mean square (RMS) roughness of the nanofilms as small as 0.13 nm that is also similar to the RMS of single crystal micrometer thick films. The $Bi_1Tm_2Fe_5O_{12}$ films demonstrate effective uniaxial anisotropy. These all make the fabricated nanofilms very promising for their potential applications in magneto-optical devices and quantum technologies.

**Introduction**

Dielectric transparent magneto-optical thin and ultrathin films with a large magneto-optical parameter are important because they can be used for integrated photonics devices [1, 2], for sensors [3-7], ultra-fast data storage [8,9], and nanophotonics, including plasmonics [10-12] and all-dielectric metasurfaces [13-16]. The films of bismuth substituted iron garnets are the most suitable for such applications. The addition of bismuth to iron garnet significantly enhances the magneto-optical interaction [35, 36]. Also, these films have low optical losses and high magnetic quality factor. The bismuth iron garnet nanofilms are of particular interest due to miniaturization of nanophotonic devices [17, 18] and the possibility of transition to quantum devices, since, for example, magnetic qubits are considered on their basis [19, 20].

There are several methods that are commonly used for fabrication of thin iron garnet films, including liquid phase epitaxy (LPE) [21-23], radio frequency (RF) magnetron sputtering [24-26], pulsed laser deposition (PLD) [27-29] and others [30-31]. LPE is a highly efficient technique for the growth of high-quality single-crystalline films with over extensive surface areas with a thickness of at least several tens of nanometers [21], but fabrication of iron garnet films with a thickness of less than 10 nm by the LPE method represents a rather challenging task due to the diffusion of substrate ions, surface roughness and nucleation effects which cause non-uniform growth and variations in film thickness leading to low overall quality. On the other hand, theRF-magnetron sputtering is applicable to nanofilm synthesis as well, though quality of the sputtered films is generally worse due to their polycrystalline character. Thus, there were grown 7-45 nm thick films fabricated by RF-sputter deposition with the Faraday effect up to 0.17 deg/µm and 0.37 deg/µm at wavelength 1550 nm for bismuth and cerium substituted yttrium iron garnet, respectively [32].

PLD thin films of iron garnet are often considered better than those fabricated using RF-sputtering. PLD provides precise control over the stoichiometry of the film through adjustment of the laser energy and deposition parameters, also generally produces films with improved crystallinity, smoother surfaces, and reduced defect densities. In this respect the PLD is considered as one of the most promising methods for growing nanometer thick iron garnet films since it allows to fabricate films of quality very close to the monocrystalline kind. For example, in [33] 8 nm thick film $Bi_1Y_2Fe_5O_{12}$ 15 nm demonstrating the Faraday rotation of -3 deg/µm at 632 nm were grown by PLD. The $Bi_3Fe_5O_{12}$ films synthesized in [34] were a bit thicker, 15 nm thick, but demonstrated very large Faraday rotation up to -4 deg/µm at a wavelength of 632 nm. Comparable results were obtained in our experiments but for $Bi_1Y_2Fe_5O_{12}$ with the lower concentration of Bi. As it has been shown in many works [35, 36], the higher the concentration of bismuth, the higher the Faraday rotation, therefore our films have a comparable magneto-optical effect.

In the current work we fabricated ultrathin nanofilms of $Bi_1Y_2Fe_5O_{12}$ with the thicknesses of 2-10 nm, with the record Faraday rotation up to 28.6 deg/µm for 2 nm thickness at a wavelength of 420 nm and 2.1 deg/µm for 2nm thickness at 632nm.

**Fabrication of samples**

The nanofilms were prepared using PLD. The target was ablated by an ArF excimer laser (2~5 Hz repetition rate) in a vacuum chamber. The laser energy was 200~250 mJ and the films deposition rate was estimated to be around 1 nm/min. The films were deposited on substrates placed on a substrate holder parallel to the target. The target to substrate distance was maintained at 4~6 cm. The base pressure of the chamber was $1\times10^{-5}$ Pa. The epitaxial film growth was attained at the substrate temperature of around 700~800 °C and at ambient oxygen pressure between 20~50 mTorr. After the deposition was completed, the film was first kept in the growth chamber for 20 minutes. After cooling, the film was transferred to the annealing furnace for annealing. Argon gas was introduced into the annealing furnace, with the temperature controlled between 800 to 850 degrees Celsius, maintained for 3 hours, followed by cooling.

Five films of bismuth-substituted yttrium iron garnet, $Bi_1Y_2Fe_5O_{12}$ (BiYIG), and bismuth-substituted thulium iron-garnet, $Bi_1Tm_2Fe_5O_{12}$ (BiTmIG), with thicknesses ranging from 2 to 10 nm were synthesized on gadolinium gallium garnet (GGG) substrates.

**Surface morphology**

The morphology of the surface of the films, the results of which are shown in Fig. 1, was studied by semi-contact atomic force microscopy (AFM) using scanning probe microscope Ntegra with modification for optical measurements (NT-MDT). The data were obtained by NSG10 series cantilevers with a needle tip of less than 10 nm, an average value of the resonant frequency in the range from 140 to 390 kHz and an average value of the force constant in the range from 3.1 to 37.6 N/m. Scanning modes and parameters were selected individually for each sample to obtain the best AFM image. The frame size during scanning was 2 x 2 µm.

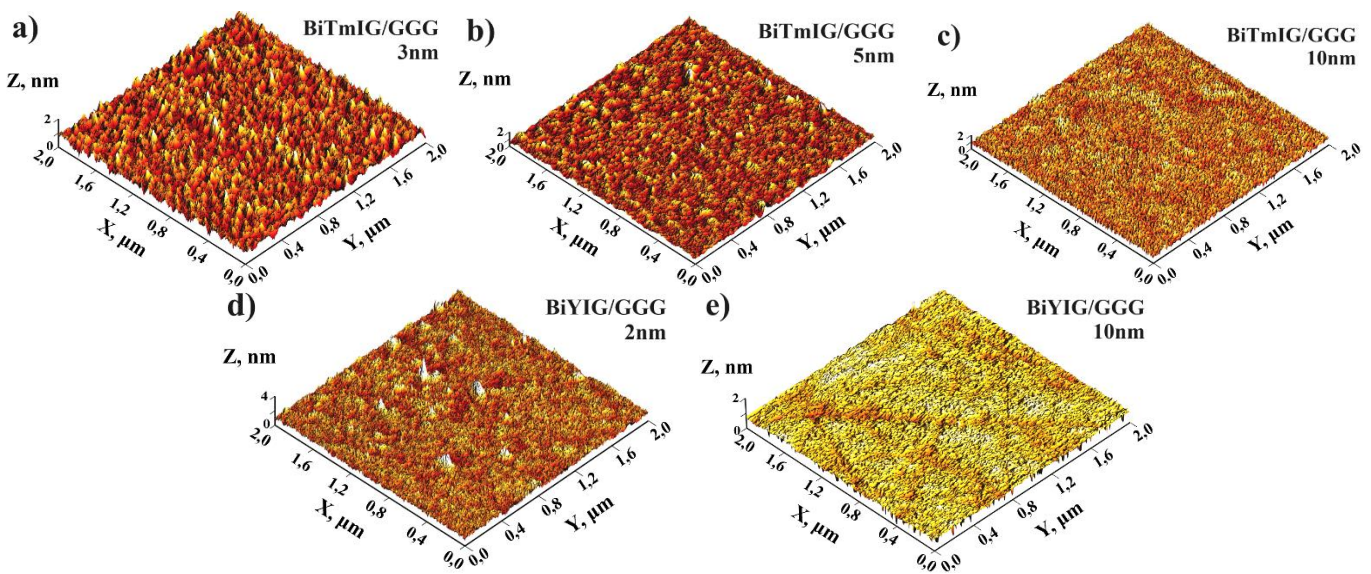

Fig. 1. AFM images of the surface of the BiTmIG and BiYIG structures. BiTmIG with thickness 3 nm (a), 5 nm (b), and 10 nm (c). BiYIG with thickness 2 nm (d) and 10 nm (e).

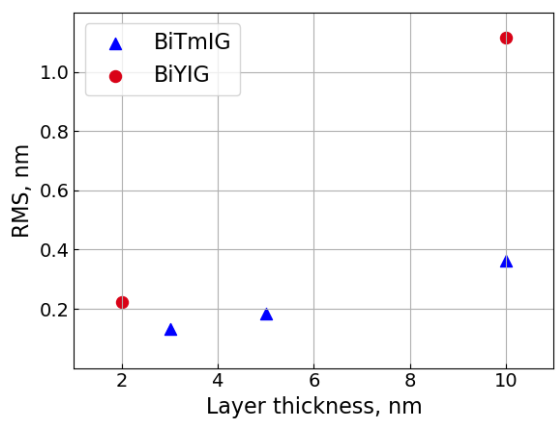

Fig. 2. Dependence of RMS roughness on the thickness of BiTmIG and BiYIG layer on GGG substrate.

Using the data from (Fig. 1) the results of the RMS roughness of five single-layer samples were obtained and presented in (Fig.2). RMS increases with increasing thickness of the BiTmIG layer: from 0.13 to 0.36 nm. This type of parameter change is also observed for other surface parameters – the height span (peak-to-peak value) and average roughness and is most likely associated with the accumulation of defects in the deposited layer. For the BiYIG layer, the roughness increases from 0.22 to 1.11 nm, which corresponds to a tendency for roughness to enhance with increasing thickness of thin films. The results obtained are in good agreement with the work [27]: an increase in the RMS value from 0.3 nm to 2.2 nm in that article is related to an increase in the thickness of the BiTmIG (using PLD technique) layer from 5 nm to 160 nm, respectively. The value of the RMS roughness is tenths of nm, these are the reference values even for epitaxial films of iron garnets, as presented in [37].

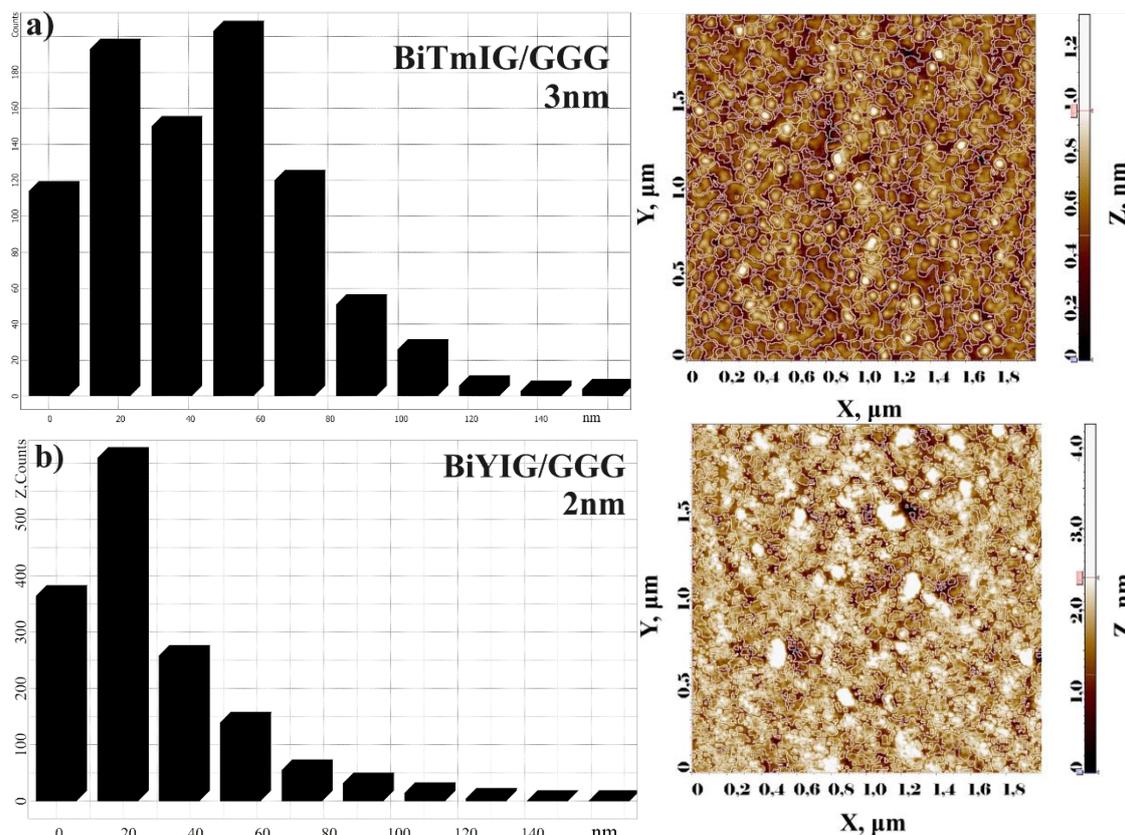

Fig. 3. A histogram showing the distribution of average grain sizes in the left column and highlighted grain boundaries in the two-dimensional AFM image in the right column of BiTmIG (3nm) (a) and BiYIG (2nm) (b) thin films.

For complete study of the morphology, a grain analysis was performed (Fig. 3). Grains are understood as inhomogeneities of the film surface. The analysis of the formed grains on the surface of single-layer films revealed the following features. The number of crystallites on the surface of BiTmIG films in a 2 x 2 µm frame increases from 870 (film thickness 3 nm) to 1070 (film thickness 5 and 10 nm). The lateral sizes of BiTmIG crystallites range from 10 to 80 nm, and larger conglomerates are also found. For a film with a thickness of 10 nm the grains become smaller and the surface has a more uniform grain size. The number of crystallites for BiYIG on the surface is about 1.5 times larger than that of BiTmIG and conglomerates consist of grains ranging in size from 10 to 50 nm with a predominance of grains of 20 nm in size. This behavior of crystallite growth can be explained by the fact that the growth of iron garnet films depends on the lattice mismatch parameter of the substrate and the film. The lower the mismatch, the easier the films grows and the higher their quality is.

**Magneto-optical properties**

Despite nanometer thickness of the films, their specific magneto-optical Faraday rotation angle is comparable to the much thicker counterparts. The spectrum of the Faraday effect $\theta_F$ in the range from 400 to 900 nm for 2-3nm and 10 nm films of BiYIG and BiTmIG is presented in (Fig. 4). The spectrum of the Faraday effect has two resonances of opposite sign. When bismuth ions are added to the garnet structure, they introduce new electronic states of the iron ions and modify the crystal field around. This alteration enhances the electro-dipole resonances within the material, leading to two distinct resonances with opposite signs in the Faraday effect spectrum [35, 36]. Also, the bismuth ions affect the magnetic properties of the iron garnets, for example, at room temperature the saturation magnetization increases [36, 38].

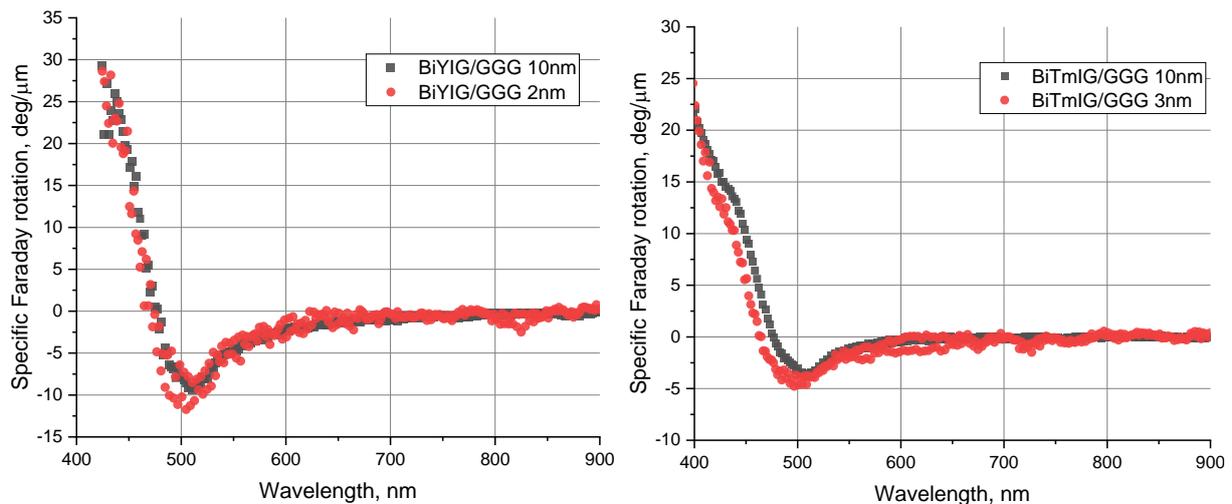

Fig. 4. Dependence of specific Faraday rotation angle spectrum of BiYIG (2 and 10nm) in the left column and BiTmIG (3 and 10 nm) in the right.

BiYIG films generally provide Faraday rotation higher than BiTmIG (Fig. 4). The highest specific Faraday rotation $\theta_F$ = 29.3 deg/µm is observed at a wavelength of 420nm for 10 nm thick BiYIG. At a conventional laser wavelength of 632 nm the Faraday rotation is -2 deg/µm, the content of Bi is only 1 ($Bi_1Y_2Fe_5O_{12}$), then an angle of about -6 deg/µm is expected for the composition of $Bi_3Fe_5O_{12}$, since the dependence is linear, as was shown in [36]. To enhance magneto-optical effects, the lattice mismatch strain should be reduced by choosing a proper composition of the film or using a different substrate. This can be attributed to the limited solubility of the Bi dopant, due to strain compensation and minimization of internal energy, as described by [39,40]. Additionally, the Faraday angle can be enhanced by increasing the Bi concentration as mentioned previously.

The films fabricated by pulsed laser deposition technique shows a bit better result that the thicker films fabricated by the liquid phase epitaxy (Fig 5). Compared to the influence of bismuth ions, lutetium and yttrium have little effect on the Faraday effect. On the other hand, to obtain an LPE film, it is necessary to match its lattice with the substrate lattice (in our case GGG). Therefore, the composition of $Bi_{0.5}Lu_{2.5}Fe_5O_{12}$ (BiLuIG) with thickness 0.35 µm was used, which is consistent with GGG and similar to fabricated by PLD $Bi_{0.5}Y_{2.5}Fe_5O_{12}$ film by Bi content [41]. The maximum Faraday rotation in this range for this film is 13.3 deg/um. There is the linear dependence between concentration of bismuth and Faraday rotation [35, 36], the higher the concentration of bismuth, the higher the Faraday rotation. For $B_1Lu_2Fe_5O_{12}$, the Faraday rotation is expected to be on the order of 26 degrees/μm, which is comparable on the nanometer PLD films.

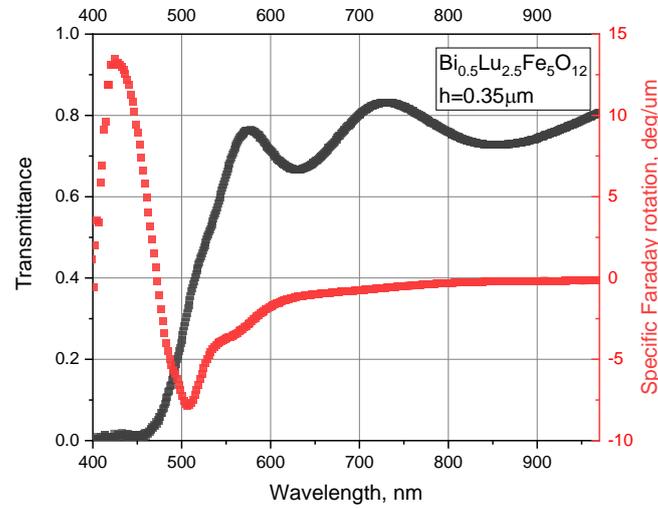

Fig. 5. Transmittance and specific Faraday rotation angle spectrum of LPE grown $Bi_{0.5}Lu_{2.5}Fe_5O_{12}$/GGG film with thickness 0.35 μm.

For the characterization of the magnetic properties of the films, the hysteresis of the magneto-optical Faraday effect was measured at 405 nm (Fig. 6). Hysteresis of the Faraday rotation angle was obtained in a normal magnetic field that varied from -0.3 T to 0.3 T using a single-channel photodetector. The laser beam of light passed through two polarizers crossed at an angle of 45 deg.

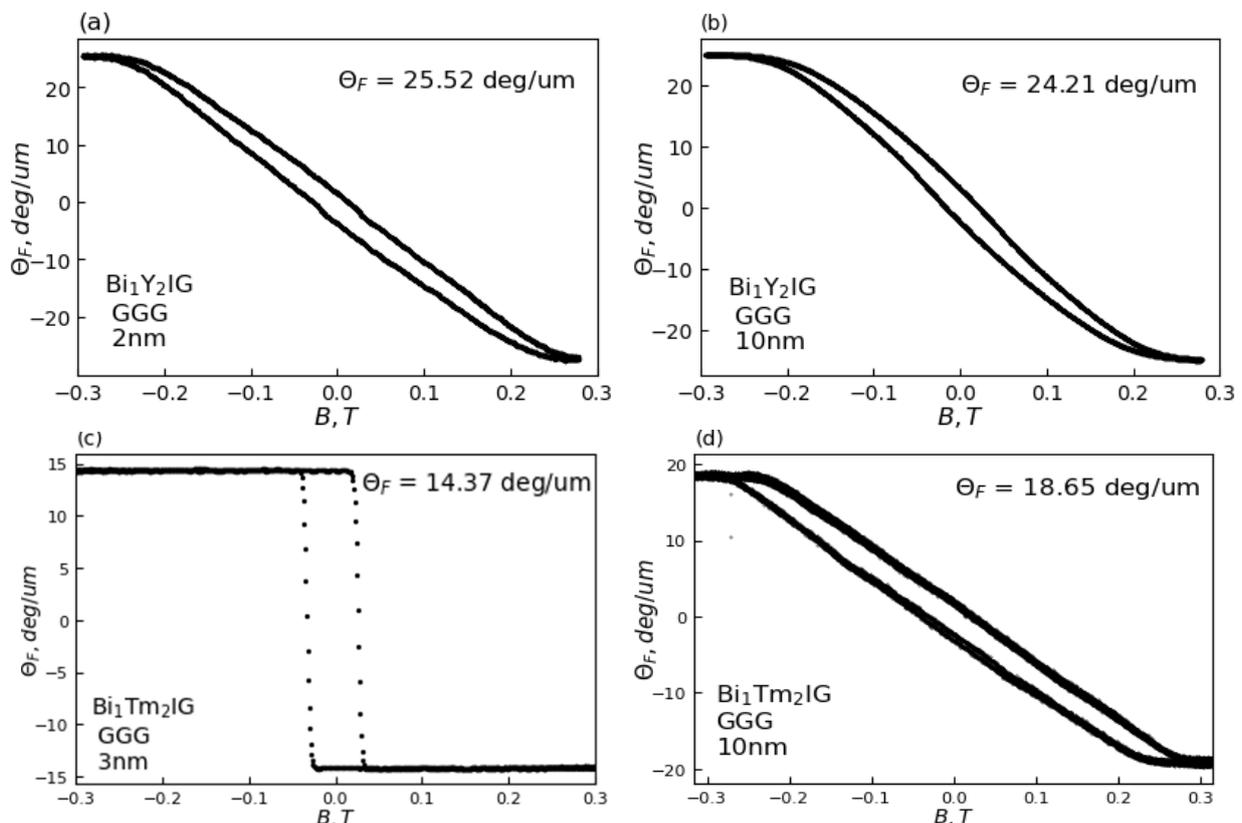

Fig. 6. The out-of-plane hysteresis of the magneto-optical Faraday effect for BiYIG: 2 nm (a) and 10 nm (b) and BiTmIG: 3 nm (c) and 10 nm (d). Laser wavelength is 405 nm, light incidence and magnetic field are along film normal.

The shape and magnitude of the hysteresis loop can reveal information about magnetic domain structure and other magnetic properties of the material. By studying the hysteresis one can find that BiYIG films have in-plane and BiTmIG films have out-of-plane effective magnetic anisotropy (taking into account influence of both uniaxial anisotropy and demagnetizing field), which has interest in ultra-fast data storage. With the increase of the film thickness effective anisotropy of BiTmIG turns from easy axis into easy plane type. The same behavior was found for yttrium iron garnet thin films [25].

The mismatch parameter between the film and substrate crystal lattices is smaller for BiTmIG than for BiYIG films. The larger ion radius of $Bi^{3+}$ (1.03 Å) compared to $Y^{3+}$ (0.9 Å) and $Tm^{3+}$ (0.869 Å) leads to an expansion of the lattice constant [24, 33]. Furthermore, the placement of $Bi^{3+}$ in the dodecahedral position has a notable impact on the growth-induced anisotropy of films. The average lattice constant of the BiYIG films is measured to be 12.459 Å, which is larger than BiTmIG lattice constant and also exceeds that of the GGG substrate, 12.383 Å.

Coercivity, saturation field, Faraday angle at wavelengths 405 nm, 420 nm (maximum positive value of the Faraday rotation for BiYIG in the range of photon energies from 1.26 to 3.17 eV), 511 nm (minimum for BiYIG and BiTmIG) are presented in Table I. The most suitable sample for the applications is one with the easy axis BiTmIG film of 3 nm thickness.

Table I. Summary of the magnetic properties of BiYIG and BiTmIG films.

| Sample | Thickness, nm | Faraday angle at 405 nm, deg/μm | Coercivity, Oe | Saturation field, Oe | Faraday angle at 420 nm, deg/μm | Faraday angle at 511 nm, deg/μm |
| --- | --- | --- | --- | --- | --- | --- |
| $Bi_1Y_2Fe_5O_{12}$ | 2 | 25.5 | 203 | 2558 | 28.6 | -9.6 |
| $Bi_1Y_2Fe_5O_{12}$ | 10 | 24.2 | 194 | 2326 | 29.3 | -9.4 |
| $Bi_1Tm_2Fe_5O_{12}$ | 3 | 14.4 | 323 | 3507 | 13.2 | -3.8 |
| $Bi_1Tm_2Fe_5O_{12}$ | 10 | 18.7 | 290 | 2820 | 16.5 | -4.3 |

**Conclusion**

We synthesized magneto-optical nanofilms of iron garnet ($Bi_1Y_2Fe_5O_{12}$ and $Bi_1Tm_2Fe_5O_{12}$) by PLD method. Notably, the film thicknesses range from 2 to 10 nm, corresponding to just a few crystal lattice constants. Reference values for the Root Mean Square (RMS) roughness in tenths of nanometers are achieved. The films exhibit relatively smooth morphology, which is important for the magneto-optical effects. The Faraday rotation spectra were measured across photon energies from 1.26 to 3.17 eV, revealing the unique magneto-optical properties of these nanometer-scale films. Moreover, thin films having effective magnetic anisotropy of the easy axis type ($Bi_1Tm_2Fe_5O_{12}$ thin film (3 nm)) are important for magnetic data storage and magnetometry. Overall, the research demonstrates the successful PLD fabrication of smooth nanofilms with excellent magneto-optical properties suitable for magneto-optical and spintronic devices.